\documentclass[12pt]{iopart}

\usepackage{graphicx}
\usepackage{epstopdf}

\usepackage{iopams}
\usepackage{footmisc}

\usepackage{xspace}
\usepackage{color}

\newcommand\etc{\textit{etc.}\xspace}
\newcommand\eg{\textit{e.g.}\xspace}
\newcommand\ie{\textit{i.e.}\xspace}

\def \Bo {B}
\def \b {\hat{\bf b}}
\def \br {b_{\psi}}

\def \kpar {\mbox{$k_{\parallel}$}}
\def \kperp {k_{\perp}}

\def \vth {\mbox{$v_{\scriptsize{\mathrm{th}}}$}}
\def \vth {v_{\scriptsize{\mathrm{T}}}}

\def \ei {q}
\def \z {z}

\newcommand{\vpar}{\ensuremath{v_{\parallel}}}
\newcommand{\vperp}{\ensuremath{v_{\perp}}}

\newcommand{\gyroavg}[1]{\left\langle#1\right\rangle_{\bf R}}
\newcommand{\angleavg}[1]{\left\langle#1\right\rangle_{\bf r}}

\providecommand\bnabla{\boldsymbol{\nabla}}

\def \rhovec { \mbox{\boldmath $\rho$} }

\def \Op {\Omega_\phi}

\newcommand{\poiss}[2]{\{#1,\;#2\}}
\newcommand{\zon}[1]{\overline{#1}}
\newcommand{\nzo}[1]{\tilde{#1}}

\begin{document}

\title{Nonlinear growth of zonal flows by secondary instability in general magnetic geometry}

\author{G G Plunk$^1$, A. Ba\~{n}\'{o}n Navarro$^2$} 
\address{$ö1$ Max-Planck-Institut f\"ur Plasmaphysik, Wendelsteinstr. 1, 17491 Greifswald, Germany} 
\address{$ö2$ Department of Physics and Astronomy, University of California, Los Angeles} 
\ead{gplunk@ipp.mpg.de} 


\begin{abstract}
We present a theory of the nonlinear growth of zonal flows in magnetized plasma turbulence, by the mechanism of secondary instability.  The theory is derived for general magnetic geometry, and is thus applicable to both tokamaks and stellarators.  The predicted growth rate is shown to compare favorably with nonlinear gyrokinetic simulations, with the error scaling as expected with the small parameter of the theory.
\end{abstract}

\maketitle

\section{Introduction}

The present work concerns the dynamics of zonal flows in magnetic confinement fusion devices, and in particular the three dimensional magnetic geometries of stellarators.  The magnetic field lines in these devices trace out nested topologically toroidal surfaces, and zonal flows are special linearly stable modes, extending across these surfaces, that suppress turbulence and thereby improve confinement \cite{diamond-zonal}.  In the absence of a nonlinear turbulent drive, they oscillate and damp, tending asymptotically to a non-zero ``residual'' level.  This linear behavior has already been studied in the stellarator context \cite{sugama-2006, sugama-2008, mishchenko, helander-mischenko}.  The nonlinear mechanism of growth, however, is less understood.  This growth can be estimated by ``secondary'' instability, and an analogous nonlinear decay mechanism is given by the ``tertiary'' instability.  The goal of the present paper is to clarify this growth mechanism, and a future paper is planned to investigate nonlinear decay.

In the theory of plasma turbulence, the notion of secondary instability, generally speaking, provides a useful tool for understanding the nonlinear processes that determine fluctuation amplitudes.  It is essentially an instability of an instability.  That is, when a linearly unstable plasma wave, called a primary instability, grows to sufficiently large amplitude, nonlinear processes arise that can drive a new mode.  The new mode, called a secondary instability, must eventually grow faster than the primary mode, since it has a growth rate proportional to the amplitude of the primary mode, and thus could ultimately cause saturation of the primary mode, by drawing energy away from it.  On the other hand, the secondary mode can also function more indirectly, by driving otherwise stable modes called zonal flows, that are involved in mode saturation by ``back-reaction'' or shearing.  This second scenario is of great importance in ion temperature gradient (ITG) turbulence in magnetic fusion devices, which is the motivation for the present work.

The secondary instability that drives zonal flows has been derived using fluid theory in two dimensions \cite{rogers-prl, strintzi-jenko}, and related instabilities have also been calculated that assume scale separation between the zonal flow and primary mode (\eg \cite{smolyakov-diamagnetic}).  Some works on zonal flow growth have modeled the effect of magnetic geometry \cite{anderson-pop, anderson-jpp}, by deriving local fluid approximations of linear theory that, however, ultimately do not account for the variation of the linear modes within a flux surface.  In the present work, we treat the ITG mode as a given parameter of the theory, so that its non-local coupling to the zonal mode may be fully calculated.  It is demonstrated here that this non-local coupling strongly affects zonal flow growth.  The derived theory is applicable to axisymmetric geometries (tokamaks) and also the fully three-dimensional geometry of stellarators.  Although the theoretical zonal flow growth rate could be computed fully analytically by using an analytical model of the ITG mode, an exact result is obtained by instead numerically simulating the ITG mode and using the result as an input of the theory; we will take this second approach here.

The outline of the paper is as follows.  In Sec.~\ref{eqn-sec} we introduce the equations and notation, and the general representation needed for three dimensional magnetic geometry.  In Sec.~\ref{deriv-sec} we present the complete details of the derivation of the secondary mode growth rate.  By directly calculating the effect of the infinite hierarchy of coupled Fourier modes, we demonstrate its equivalence to the truncated ``four-wave'' approximation, agreeing with expectations from previous work \cite{strintzi-jenko}.  Because the derivation is asymptotically exact, it should agree with gyrokinetic simulations in the limit that the scale of the modes exceeds the ion Larmor radius.  Therefore, in Sec.~\ref{numeric-sec} we test this, and find that the difference between the theoretical and numerical growth rates indeed scales with the small parameter of the theory.  Although the theory is valid for stellarator geometries, we choose for the sake of simplicity to make this first numerical demonstration in circular tokamak geometry.  In Sec.~\ref{conclusion-sec} we summarize and make some final remarks.

\section{Equations}\label{eqn-sec}

Let us consider the nonlinear electrostatic ion gyrokinetic equation,

\begin{equation}
\frac{\partial h}{\partial t} +  \vpar \frac{\partial}{\partial \z} h + ({\bf v}_E + {\bf v}_d) \cdot \bnabla_\perp h + {\bf v}_E \cdot \bnabla F_0 = \frac{\ei}{T_i}\frac{\partial \gyroavg{\phi}}{\partial t} F_0,\label{gk-eqn}
\end{equation}

\noindent where $h({\bf R}_{\perp}, \z, \varepsilon, \mu)$ is the non-adiabatic part of the ion distribution function, which depends on the gyro-center coordinate ${\bf R}_\perp$ and the independent velocity coordinates are $\mu = m_i \vperp^2/(2\Bo)$ and $\varepsilon = m_i v^2/2 =Êm_i(\vpar^2 + \vperp^2)/2$ where $m_i$ is the ion mass.  We employ the following mostly standard conventions:  ${\bf v}_E = \b \times\bnabla\gyroavg{\phi}/\Bo$ is the gyro-averaged $E\times B$ velocity; ${\bf v}_d = \Omega_c^{-1}\b\times(\vpar^2 \b\cdot\bnabla\b + \vperp^2/2 \bnabla \ln \Bo)$ is the magnetic drift velocity, ${\bf \Bo} = \b \Bo$ is the equilibrium magnetic field, $\Omega_c = \ei \Bo/m_i$ is the ion cyclotron frequency; $\ei = Z e$ is the charge of an ion; $T_i$ and $T_e$ are the background ion and electron temperatures; $F_0 = n_0 (\pi \vth^2)^{-3/2}\exp(-v^2/\vth^2)$ is the background Maxwellian distribution function, where the ion thermal velocity is $\vth = \sqrt{2 T_i/m_i}$.  The gyro-average is defined

\begin{equation}
\gyroavg{A({\bf r})} = \frac{1}{2\pi}\int_0^{2\pi} d\vartheta A({\bf R} + \rhovec(\vartheta)).
\label{gyro-avg-def}
\end{equation}

\noindent where $\rhovec(\vartheta) = \b\times{\bf v}/\Omega_c$ and ${\bf v} = \vpar \b + \vperp[\hat{\bf e}_1 \cos(\vartheta) + \hat{\bf e}_2 \sin(\vartheta)]$, and $\hat{\bf e}_1\times\hat{\bf e}_2 = \b$.  The electrostatic potential $\phi$ is determined by the quasi-neutrality constraint, assuming adiabatic electrons:

\begin{equation}
\frac{1}{n_0}\int d^3{\bf v} \left<h\right>_{\bf r} = \frac{q \phi}{T_i} + \frac{e\nzo{\phi}}{T_e}.\label{qn-eqn}
\end{equation}

\noindent Here, the non-zonal electrostatic potential is denoted $\nzo{\phi} = \phi - \zon{\phi}$ where the zonal average is defined in Eqn.~\ref{zon-def} below, and the angle average at particle position ${\bf r}$ is given by

\begin{equation}
\angleavg{A({\bf R})} = \frac{1}{2\pi}\int_0^{2\pi} d\vartheta A({\bf r} - \rhovec(\vartheta)).
\label{angle-avg-def}
\end{equation}

\noindent Note that for simplicity trapped electrons have been neglected; the modified Boltzmann electron model is formally only valid in this limit.  

\subsection{Representing turbulence in general magnetic geometry}

In order to keep the theory as general as possible, we would like to allow a representation that is slightly more general than the usual flux Fourier series applied in flux tube geometry.  The eikonal representation is used \cite{roberts-taylor}, \ie the fluctuations are decomposed as a sum of modes, as follows:

\begin{equation}
\phi = \sum_{S} \hat{\phi}_S \exp(i S), \label{eik-phi} \\
\end{equation}

\noindent The coefficients $\hat{\phi}_S = \hat{\phi}_S(\psi, \alpha, \z)$, vary slowly in space, where $\z$ denotes the coordinate of distance along the field line, $2\pi\psi$ is the toroidal magnetic flux and $\alpha$ is the final flux coordinate such that the equilibrium magnetic field is written ${\bf \Bo} = \bnabla\psi\times\bnabla\alpha$.  The eikonal functions $S = S(\alpha, \psi)$ also vary slowly in space, but the local perpendicular wavenumber ${\bf k}_{\perp} = \bnabla S = k_\psi \bnabla \psi + k_\alpha \bnabla \alpha$ is large, \ie ${\bf k}_\perp \sim {\cal O}(1/\rho)$, so that the factor $\exp(i S)$ is a rapidly varying phase factor.  The sum in Eqn.~\ref{eik-phi} can equivalently be written as a sum over the functions $k_\psi = \partial S/\partial \psi$ and $k_\alpha = \partial S/\partial \alpha$, \ie.

\begin{equation}
\phi = \sum_{k_\psi, k_\alpha} \hat{\phi}_{\bf k} \exp(i k_\psi \psi + i k_\alpha \alpha).\label{phi-sum-rep}
\end{equation}

\noindent Note that scale separation, \ie $\kperp \sim 1/\rho_i \gg L$, where $L$ is the variation scale of the equilibrium, ensures these forms are equivalent since $\bnabla_\perp \phi \approx \sum_{\bf k} i {\bf k}_{\perp} \hat{\phi}_{\bf k}$.  It also implies that the wavenumbers $k_\alpha$ and $k_\psi$ are real, because otherwise $\phi$ would grow exponentially over a macroscopic distance and quickly violate the assumption of small fluctuations that underlies gyrokinetics, $q\phi/T_i \ll 1$.  Thus, Eqn.~\ref{phi-sum-rep} constitutes a local Fourier series, but the point we emphasize here is that the waveneumbers $k_\alpha$ and $k_\psi$ are free to also vary slowly in the $\alpha$ and $\psi$ coordinates.  This fact can matter when considering turbulence in larger domains such as a flux surface or a full toroidal volume.

The flux surface average is (see \eg \cite{helander-review})

\begin{eqnarray}\label{zon-def}
\zon{\phi} &\equiv \frac{1}{V'}\int_0^{2\pi} d\alpha \int \frac{d\z}{\Bo} \phi(\psi, \alpha, \z) \\
&= \frac{1}{V'}\sum_{k_\psi, k_\alpha} \int_0^{2\pi} d\alpha \int \frac{d\z}{\Bo}\exp (i k_\psi \psi + i k_\alpha \alpha)\hat{\phi}_{\bf k}\\
&= \frac{1}{V'}\sum_{k_\psi, k_\alpha = 0} \int_0^{2\pi} d\alpha \int \frac{d\z}{\Bo}\exp (i k_\psi \psi) \hat{\phi}_{\bf k},
\end{eqnarray}

\noindent where the integral in $\z$ is taken over one toroidal transit, and $V'(\psi) = \int d\alpha \int d\z/\Bo(\psi,\alpha,\z)$.  Note that we have used the fact that, due to scale separation, the non-zero $k_\alpha$ terms will average to zero upon integration over $\alpha$.  The integration over $\alpha$ still appears due to the fact that $\hat{\phi}$, $\Bo$, \etc vary (smoothly) in $\alpha$.


\section{Derivation of the global secondary instability in general geometry}\label{deriv-sec}

The secondary instability of Rogers and Dorland \cite{rogers-prl} is a nonlinear mechanism for the generation of zonal flows by ion-scale turbulence.  Two key assumptions make it possible to obtain an analytic result for this mode: (1) the Larmor scale is small $\kperp^2\rho_i^2 \ll 1$, and (2) the ion motion parallel to the magnetic field is slow, $\kpar\vth \ll \omega$.  The second assumption is called called ``quasi-two-dimensional'' (or ``pseudo-three-dimensional'' \cite{hasegawa-mima}) because parallel electron motion is retained, but the ion dynamics is only two-dimensional \cite{dorland-thesis, cohen}.  That is, the two-dimensional motion of the ions is calculated in a three-dimensional domain, \ie a magnetic geometry with shear, {\em etc}, with coupling along the field line being provided by the electrons, zeroing out the flux-surface-average density fluctuations.  For the secondary instability that drives zonal flows, it is thus important to include information about the 3D mode structure of the primary mode.  This is justified in the strongly ballooning limit, in which case parallel ion motion is slow, but nevertheless determines the the structure of the mode along the field line, \ie its ``ballooning structure'' (see \eg \cite{plunk-itg}).  In this limit, the ballooning structure does not strongly affect the primary mode growth, but we will see that it does strongly affect the secondary instability.  This is essentially because the global secondary instability has a strong zonal component, and it is therefore a non-local instability that depends on the properties of the turbulence across the entire flux surface.  Intuitively, a zonal mode is driven more effectively by fluctuations that occupy a large fraction of a flux surface.  This idea has been mentioned by previous authors \cite{li-kishimoto, citrin-pop-2012}, but has not been investigated in detail.

We will now derive an expression for the global secondary instability.  We first neglect the linear terms from the gyrokinetic equation.  Then, assuming $\kperp^2\rho_i^2 \ll 1$, we may expand the Bessel functions and derive a fluid system (this is a 3D version of the zeroth-order system of \cite{plunk-njp}).  Note that the fluid system is derived as an exact asymptotic limit of the gyrokinetic system, \ie there is no truncation performed in the fluid moment hierarchy.  The system is

\begin{eqnarray}
\frac{\partial  \nzo{\phi}}{\partial t} + \poiss{\zon{\phi}}{\nzo{\phi}} = 0,\label{nzo-phi-eqn-0}\\
\frac{\partial}{\partial t}(-\zon{\rho^2\nabla_{\perp}^2 \phi}) + \zon{\poiss{\phi}{-\rho^2\nabla_{\perp}^2\phi - \rho^2\nabla_{\perp}^2\chi}}\nonumber\\
+ \zon{\poiss{\rho^2\nabla_{\perp}^2 \phi}{\chi}} + \zon{\rho^2\nabla_{\perp}^2\poiss{\phi}{\chi}} = 0,\label{zon-phi-eqn-0}\\
\frac{\partial \chi}{\partial t} + \poiss{\phi}{\chi} = 0,\label{chi-eqn-0}
\end{eqnarray}

\noindent where we recall that $\nzo{\phi} = \phi - \zon{\phi}$, and note that the thermal Larmor radius varies in space, \ie 

\begin{equation}
\rho = \frac{m\vth}{\sqrt{2}q \Bo(\psi, \alpha, \z)},
\end{equation}

\noindent and $\chi$ (proportional to the ion gyrocenter perpendicular temperature) is defined

\begin{equation}
\chi({\bf r}) = \frac{1}{q n_0}\int d^3{\bf v} \left(\frac{m\vperp^2}{4}\right) \left(h({\bf r}) - \frac{q\phi({\bf r})}{T_i} f_M\right).
\end{equation}

\noindent For arbitrary functions $f$ and $g$, the Poisson bracket is defined

\begin{equation}
\poiss{f}{g} = \frac{1}{\Bo} (\b\times\bnabla f )\cdot \bnabla g.
\end{equation}

We consider the primary mode as a given solution of the linearized gyrokinetic equation, and denote this solution by $\phi_p$, $\chi_p$.  We emphasize that the $z$-dependence of the primary mode is determined by parallel ion motion, which therefore must be retained in the linearized gyrokinetic equation, but can be neglected in the system above, which is derived to describe the secondary instability.  The primary mode has the form

\begin{eqnarray}
\phi_p =  \phi_{p0} \exp(i S_p) + \mathrm{c.c.},\label{phi-p-def}\\
\chi_{p} = \chi_{p0} \exp(i S_p) + \mathrm{c.c.},\label{chi-p-def}
\end{eqnarray}

\noindent were c.c. means ``complex conjugate,'' $\chi_{p0}(\psi, \alpha, \z)$ and $\phi_{p0}(\psi, \alpha, \z)$ are complex functions.  The zonal average of the primary is assumed to be zero $\partial S_p/\partial \alpha \neq 0$, \ie it has no component with $k_\alpha = 0$.  In axisymmetric geometry, this assumption is trivial because $k_\alpha$ is everywhere the same, \ie all field lines are the same.  In non-axisymmetric geometry, $k_\alpha$ of a global mode is free to vary slowly over a surface, but we will assume it is nowhere zero.  This is justified because a mode is locally stable wherever $k_\alpha = 0$, contradicting the assumption that it is globally unstable.

We will now assume that the linear mode is sufficiently localized (in $\alpha$ and $\z$) such that it does not significantly interact with itself nonlinearly (as modes that are very extended in ballooning space can do).  For a global eigenmode (a mode that is everywhere on the surface evolving with the same complex frequency), this means that the mode structure in physical space is close to a single eikonal, \ie the structure is the same in ballooning (the ``covering space'' \cite{dewar-glasser}) and physical space.  Alternatively it may be convenient to not require the primary mode to be a global mode of the system, but rather a linearly growing structure that is locally wave-like (\ie a wave packet).  In either case, this localization will enable us to focus on the secondary mode as a saturation mechanism, and rule out nonlinear self-interaction.

The secondary mode is denoted by $\phi_s$ and $\chi_s$, where $\phi_s \ll \phi_p$ and $\chi_s \ll \chi_p$.  Now taking $\phi = \phi_p + \phi_s$ and $\chi = \chi_p + \chi_s$ and expanding the nonlinearity, Eqns.~\ref{nzo-phi-eqn-0}-\ref{chi-eqn-0} yield

\begin{eqnarray}
\frac{\partial  \nzo{\phi}_s}{\partial t} + \poiss{\zon{\phi}_s}{\nzo{\phi}_p} = 0,\label{nzo-phi-eqn-s}\\
\frac{\partial}{\partial t}(-\zon{\rho^2\nabla_{\perp}^2 \phi}_s) + \zon{\poiss{\phi_p}{-\rho^2\nabla_{\perp}^2\phi_s - \rho^2\nabla_{\perp}^2\chi_s}} + \zon{\poiss{\rho^2\nabla_{\perp}^2 \phi_p}{\chi_s}} \nonumber\\ 
+ \zon{\poiss{\phi_s}{-\rho^2\nabla_{\perp}^2\phi_p - \rho^2\nabla_{\perp}^2\chi_p}} + \zon{\rho^2\nabla_{\perp}^2\poiss{\phi_p}{\chi_s}}   \nonumber\\ 
+ \zon{\poiss{\rho^2\nabla_{\perp}^2 \phi_s}{\chi_p}} + \zon{\rho^2\nabla_{\perp}^2\poiss{\phi_s}{\chi_p}} = 0,\label{zon-phi-eqn-s}\\
\frac{\partial \chi_s}{\partial t} + \poiss{\phi_p}{\chi_s} + \poiss{\phi_s}{\chi_p} = 0.\label{chi-eqn-s}
\end{eqnarray}

\noindent The above differential system is linear in $\phi_s$ and $\chi_s$ with coefficients that are periodic (in their fast spatial variation) due to the dependence $\exp(i S_p)$, and so Floquet theory applies.  Despite the smooth three-dimensional variation that is present, the fast variation of the coefficients is only in one dimension, namely in the direction of $\bnabla S_p$.  The direction that is binormal to $\bnabla S_p$ and $\b$ can be considered ignorable, so we may assume a simple wave-like dependence, \ie $\exp(i \lambda)$, such that $\b\cdot\bnabla\lambda = \bnabla S_p \cdot \bnabla \lambda = 0$.  We have another free eikonal function corresponding to the Floquet exponent, which we will denote $\nu(\psi, \alpha)$; its gradient is parallel to that of $S_p$.  In order to ensure that the secondary mode has a non-zero zonal component, we would like the overall factor $\exp(i \lambda + i \nu)$ to have a non-zero zonal average.  This is ensured by requiring $\partial \nu/\partial\alpha = -\partial \lambda/\partial\alpha$.  Thus we combine these eikonal functions into a single function $\bar{S}(\psi) = \lambda + \nu$.

We may now formally write the secondary mode in terms of a series in harmonics of $\exp(i S_p)$:

\begin{eqnarray}
\phi_s = \exp(i \bar{S} - i \omega t) \sum_{n} c_n \exp(i n S_p),\label{phi-s-def}\\
\chi_{s} = \exp(i \bar{S} - i \omega t) \sum_{n} d_n \exp(i n S_p),\label{chi-s-def}
\end{eqnarray}

\noindent where we note that $\bar{S} = \bar{S}(\psi)$, and the $c_n$ and $d_n$ are functions of $\z$, $\psi$ and $\alpha$.  The secondary mode $\phi_s$ of Eqn.~\ref{phi-s-def} can be split into zonal and non-zonal components

\begin{eqnarray}
\zon{\phi}_s = \zon{c}_0 \exp(i \bar{S} - i \omega t),\label{zon-sec-mode-def}\\
\nzo{\phi}_s = \exp(i \bar{S} - i \omega t)\left[\nzo{c}_0 + \sum_{n \neq 0}c_n \exp(i n S_p) \right],\label{nzo-sec-mode-def}
\end{eqnarray}

\noindent whereas the primary mode has no zonal component, $\phi_p = \nzo{\phi}_p$.  Now, substituting Eqns.~\ref{phi-p-def}, \ref{zon-sec-mode-def}, and \ref{nzo-sec-mode-def} into Eqn.~\ref{nzo-phi-eqn-s}, we obtain an equation for each component of $\nzo{\phi}_s$

\begin{eqnarray}
\nzo{c}_0 = 0,\\
c_{1} = i \frac{\Op}{\omega} j \zon{c}_0,\label{cpd-eqn}\\
c_{- 1} = -i \frac{\Op}{\omega} j^* \zon{c}_0,\label{cmd-eqn}\\
c_{n} = 0,\quad |n| \geq 2,
\end{eqnarray}

\noindent where $j = \phi_{p0}/|\phi_{p0}|$, and

\begin{equation}
\Op \equiv \frac{(\b\times\bnabla\bar{S})\cdot\bnabla S_p}{\Bo} |\phi_{p0}| = \frac{\partial S_p}{\partial \alpha}\frac{\partial \bar{S}}{\partial \psi} |\phi_{p0}|,\label{Op-def}
\end{equation}

\noindent Thus, only three components of $\phi_s$, namely the principal component $\zon{c}_0$ and the two sidebands $c_{\pm 1}$, are needed for the calculation.  For the temperature component of the secondary, Eqn.~\ref{chi-s-def}, the series does not truncate by itself, and so we must retain the infinite hierarchy of coefficients $d_n$.  Fortunately, these may be computed using continued fractions; for brevity, we put this part of the calculation in \ref{d-appx}.

Finally we turn to Eqn.~\ref{zon-phi-eqn-s}, which will yield the dispersion relation.  Note that the first term may be simplified because $\nzo{c}_0 = 0$:

\begin{eqnarray}
\zon{-\rho^2\nabla_{\perp}^2 \phi_s} &= -\zon{c}_0 \left(\frac{\partial \bar{S}}{\partial \psi}\right)^2\zon{\rho^2 |\bnabla\psi|^2} \exp(i \bar{S})\nonumber\\
&=-\zon{\br} \zon{c}_0 \exp(i \bar{S})
\end{eqnarray}

\noindent where we define 

\begin{equation}
\br = \rho^2 \left(\frac{\partial \bar{S}}{\partial \psi}\right)^2 |\bnabla\psi|^2.
\end{equation}

\noindent The remaining terms of Eqn.~\ref{zon-phi-eqn-s} consist of coupling terms between the sidebands, $c_{\pm 1}$ and $d_{\pm 1}$, and the primary mode.  Manipulating these terms, Eqn.~\ref{zon-phi-eqn-s} yields the following 

\begin{eqnarray}
\omega \bar{c}_0 \zon{\br} =  &2 i\zon{\Op b \left[(r^*+j^*)c_{1} + (r + j)c_{-1} + j^* d_{1} + j d_{-1}\right]} \nonumber\\ &+ i\zon{\Op \br \left[2 j^* d_{1} - 2 j d_{-1} + j c_{1} - j^* c_{-1}\right]},\label{zon-phi-eqn-s-2}
\end{eqnarray}

\noindent where $j = \phi_{p0}/|\phi_{p0}|$, $r = \chi_{p0}/|\phi_{p0}|$, and

\begin{equation}
b = \rho^2 \Bo \frac{\partial S_p}{\partial \alpha}\frac{\partial \bar{S}}{\partial \psi}.
\end{equation}

\noindent We now substitute the expressions for $d_{\pm 1}$ (Eqns.~\ref{dpd-eqn}-\ref{dmd-eqn}) and $c_{\pm 1}$ (Eqns.~\ref{cpd-eqn}-\ref{cmd-eqn}) into Eqn.~\ref{zon-phi-eqn-s-2}.  All of the resulting terms are then proportional to $\zon{c}_0$, and this constant can be divided out, leaving the dispersion relation for the mode.  After some algebra, we find the simple result

\begin{equation}
\omega^2\zon{\br} = -2\zon{\Op^2 \br \left(1 + \frac{\chi_{p0}}{\phi_{p0}} + \frac{\chi_{p0}^*}{\phi_{p0}^*} \right)},\label{full-result-eqn}
\end{equation}

\noindent where $\Op$ is defined by Eqn.~\ref{Op-def}.  Note that the radial wavenumber of the primary $\partial S_p/\partial \psi$ does not enter explicitly into the final result.  It is also interesting to consider that when we can neglect the spatial variation of the magnetic field (\ie $|\bnabla \psi|$, $\Bo$, $\partial S_p/\partial \alpha$, \etc) the only affect of the geometry on the secondary mode is that the growth rate is determined by the RMS average of the primary mode on the flux surface.  Thus, to induce similar secondary (zonal flow) growth rates, a localized mode must grow to a higher peak amplitude than a mode than is more uniformly distributed on a flux surface.

Remarkably, one can obtain the same dispersion relation as Eqn.~\ref{full-result-eqn} by performing the truncation $d_n = 0$ for $|n| \geq 2$, which is commonly done to obtain an analytic result.  This explains why such truncation has been found to be an excellent approximation to the full-series numerical gyrokinetic solutions (see \eg \cite{plunk-pop-07}).  Note that the unsheared slab result is obtained by taking the functions $\Op$, $\br$, $\phi_{p0}$, and $\chi_{p0}$ to be independent of $\alpha$ and $\z$.  The resulting expression can be compared with Eqn.~(D43) of Ref.~\cite{plunk-njp} (revealing a sign error in the earlier result).

\subsection{Numerical simulation of zonal flow growth by secondary instability} \label{numeric-sec}

Because Eqn.~\ref{full-result-eqn} is an exact asymptotic result, it should agree with fully gyrokinetic simulations, in arbitrary geometry, in the applicable limit, $\kperp^2\rho^2 \ll 1$.  We test this using the GENE code \cite{jenko-dorland-pop}, using a concentric circular model tokamak geometry (``$\hat{s}$-$\alpha$'' geometry, without shift $\alpha = 0$).  This is a significant simplification, as compared with general stellarator geometry, but is a good initial test because it allows for nontrivial variation in the primary mode eigenfunctions along the field line.  The parameter $\hat{s}$, representing the global shear in the magnetic field, is varied, along with the background temperature gradient, and the primary mode wavenumber.  We can thereby sample parameter space to confirm that theory is able to uniformly describe different physical regimes.

For the model geometry, $|\bnabla \psi|$ is constant (independent of $\z$), which somewhat simplifies the evaluation of $\br$ in Eqn.~\ref{full-result-eqn}.  The expression can then be straightforwardly translated into simulation units and evaluated directly.  

We perform direct numerical simulations of both the primary and secondary instabilities.  The primary mode solution is used as input for the theoretical expression of Eqn.~\ref{full-result-eqn}.  The theoretical growth rate obtained can then be compared directly with the numerically observed growth rate, obtained by directly simulating the secondary mode, and fitting an exponential to the time dependence the evolving zonal potential $\zon{\phi}(t)$.

The secondary mode is simulated as follows.  The primary mode is artificially frozen ($\partial/\partial t \rightarrow 0$), and the linear terms for the secondary mode are also removed, so its dynamics are fully determined by the nonlinear term.  As long as parallel ion motion remains negligible (as assumed here) this is equivalent to a simulation without these modifications in the limit of large primary mode amplitude; note that the finite-$\kpar$ secondary mode (see \cite{cowley-kulsrud}) has no zonal component, and is outside the scope of our work.  Critically, our procedure avoids the need for tedious tuning of the simulation parameters to obtain a converged result.  Another advantage of removing the linear terms, both theoretically and numerically, is that the solution obtained is proportional to the amplitude of the primary mode, so can be normalized to yield an easily reproducible result with a clear interpretation.  Because very few wavenumbers are required for the mode, the (nonlinear) simulation of the secondary mode is not much more computationally expensive than the (linear) simulation of the primary mode.

\begin{figure}
\includegraphics[width=0.95\columnwidth]{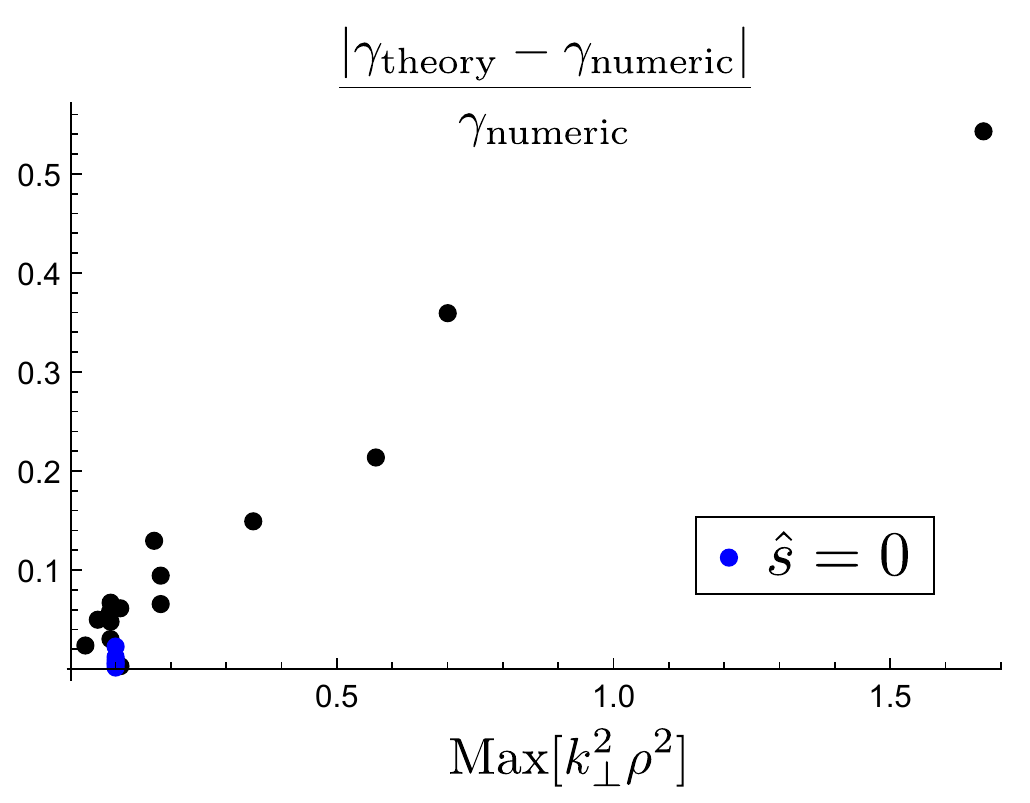}
\caption{Relative difference between the theoretical (Eqn.~\ref{full-result-eqn}) and numerical zonal flow growth rates, as a function of the small parameter $\kperp^2\rho^2$.  Points corresponding to zero magnetic shear are blue; other points correspond to finite magnetic shear.}
\label{RD-theory-vs-GENE-fig}
\end{figure}

We considered several values of $\hat{s}$ in the range of $0$ to $1$, several temperature gradients $R/L_T$ in the range of $10$-$20$, and several wavenumbers with $k_x \rho$ and $k_y \rho$ in the range of $0.05$-$0.3$.  The variation of the primary mode along the field line is sensitive to magnetic shear, allowing us to test the new prediction of theory, namely that the growth rate is sensitive to this variation.  The set of parameters was also chosen to verify that the error, as measured by the difference between the theoretical and numerical growth rates, depends mainly on the size of the small parameter $\kperp^2\rho^2$.  This comparison is complicated by the fact that the small parameter of the theory in fact varies along the field line in the presence of finite magnetic shear.  We thus chose to measure the largest value of $\kperp$ as a representative value.  As demonstrated by Fig.~\ref{RD-theory-vs-GENE-fig}, we find agreement between the numerical and theoretical results, with corrections that scale linearly with the small parameter, as expected.

\section{Conclusion}\label{conclusion-sec}

We have presented a theory of nonlinear zonal flow growth in general magnetic geometry, and validated it using nonlinear gyrokinetic simulations.  The theory is valid in the electrostatic limit, and assumes adiabatic electrons.  It applies to both axisymmetric (tokamak) and non-axisymmetric (stellarator) geometries.  We note that even in the tokamak context, the role of magnetic geometry has never previously been accounted for, and so the result may be applicable to well-studied tokamak regimes.  However, it will be especially interesting to apply the theory to the complex three dimensional geometries of stellarators, or tokamak-related problems with ``extreme'' magnetic geometry, \eg in the edge where local magnetic shear is strong.

In the solution, given by Eqn.~\ref{full-result-eqn}, magnetic geometry appears in two places.  First, it enters through the structure of the primary mode, \ie the variation of the electrostatic potential and temperature perturbation across the magnetic flux surface.  Second, it enters through the average of the radial wavenumber, which depends on the relative distribution of flux surface compression on the surface.  The first of these effects could be estimated by analytical models of linear instability, but can also be computed numerically using a gyrokinetic code, fully accounting for magnetic geometry, in local or global settings.  We note also, that the choice of what model to use for the primary mode ultimately depends on what linear physics is believed to govern the nonlinear state, and therefore a ``more complete'' primary mode calculation may not actually be a better choice to try to understand the turbulence.

Finally, we would like to reiterate the point that the nonlinear growth mechanism is one process among several that determines the role of zonal flows in turbulence.  That is, one must generally also take into account linear decay mechanisms (both collisionless and collisional) and nonlinear decay mechanisms.  However, combining an understanding of those decay mechanisms with an estimate for growth, such as the one presented here, we believe it may be feasible, with the help of only linear gyrokinetic simulations, to model the full nonlinear response of zonal flows in fusion experiments with complex magnetic geometries.

\section{Acknowledgements}

The authors wish to thank T. Bird for discussions that initiated this work, and T. G{\"o}rler and M. J. Pueschel, for help with the numerical calculation of secondary modes.  This research has received funding from the European Research Council under the European Unions Seventh Framework Programme (FP7/2007V2013)/ERC Grant Agreement No. 277870. The gyrokinetic simulations presented in this work used resources of the National Energy Research Scientific Computing Center, a DOE Office of Science User Facility supported by the Office of Science of the U.S. Department of Energy under Contract No. DEAC02-05CH11231.

\appendix
\section{Calculation of $\chi_s$}\label{d-appx}

Eqn.~\ref{chi-eqn-s} leads to expressions for the $d_n$.  Using the fact that $c_n = 0$ for $|n| \geq 2$ we write these as

\begin{eqnarray}
&-i \zeta d_0 = \left(c_{-1}r_{+} - c_{1} r_{-} + d_1 j_{-} - d_{-1} j_{+}\right),\label{d-0-eqn}\\
&-i \zeta d_{\pm 1} = \pm \left(\bar{c}_0 r_{\pm} + d_{\pm 2} j_{\mp} - d_0 j_{\pm}\right),\label{d-1-eqn}\\
&-i \zeta d_{\pm 2} = \pm \left(c_{\pm1} r_{\pm} + d_{\pm 3} j_{\mp} - d_{\pm 1} j_{\pm}\right),\label{d-2-eqn}\\
&-i\zeta d_n = d_{n+1} j_{-} - d_{n-1} j_{+},\quad |n| \geq 3\label{d-large-n-eqn}
\end{eqnarray}

\noindent where we define $j = \phi_{p0}/|\phi_{p0}|$, $ j_{+} = j$,  $j_{-} = j^*$, $r = \chi_{p 0}/|\phi_{p 0}|$, $r_{+} = r$, and $r_{-} = r^*$, and 

\begin{equation}
\zeta = \omega/\Op.
\end{equation}

\noindent We solve Eqn.~\ref{d-large-n-eqn} for positive and negative $n$ separately.  For positive $n \geq 3$ we define $q_n = j^* d_n/d_{n-1}$ and find

\begin{equation}
q_n = \frac{1}{i\zeta + q_{n+1}}.
\end{equation}

\noindent Iterating this we obtain

\begin{equation}
q_n =
\frac{1}{i\zeta + 
\frac{1}{i\zeta +
\frac{1}{i\zeta + ...
}}}.
\end{equation}

\noindent This is a special continued fraction $x$ of the form $x = 1/(y+x)$, which has the solution $x = (-y \pm \sqrt{y^2 + 4})/2$, \ie

\begin{equation}
q_n = \frac{1}{2}\left(-i\zeta \pm \sqrt{4-\zeta^2}\right).\label{qn-soln}
\end{equation}

\noindent Now for $n \leq -3$ we define $s_n = j d_n/d_{n+1}$ and similarly find

\begin{equation}
s_n =
\frac{1}{-i\zeta + 
\frac{1}{-i\zeta +
\frac{1}{-i\zeta + ...
}}},
\end{equation}

\noindent which implies

\begin{equation}
s_n = \frac{1}{2}\left(i\zeta \pm \sqrt{4-\zeta^2}\right).\label{sn-soln}
\end{equation}

\noindent The signs in the solutions \ref{qn-soln} and \ref{sn-soln} for $q_n$ and $s_n$ can be determined by the condition $|q_n| < 1$ and $|s_n| < 1$, which is required for the convergence of the series in Eqn.~\ref{chi-s-def}.  This leads to the result

\begin{equation}
d_{\pm 3} = \mp j_{\mp} d_{\pm 2}s,\label{d3-soln}
\end{equation}

\noindent where $s$ is defined as 

\begin{equation}
s = \frac{i\zeta}{2} \pm \sqrt{1 - \zeta^2/4}\label{s-def},
\end{equation}

\noindent with the sign chosen such that $|s| < 1$.  We may now combine Eqn.~\ref{d3-soln} with Eqns.~\ref{d-0-eqn}-\ref{d-2-eqn} to obtain the following expressions for $d_{\pm 1}$:

\begin{eqnarray}
&d_{1} = i \bar{c}_0\frac{\Op}{\omega} r ,\label{dpd-eqn}\\
&d_{-1} = - i \bar{c}_0\frac{\Op}{\omega} r^*.\label{dmd-eqn}
\end{eqnarray}

\noindent It is worth noting that the above expressions for $d_{\pm 1}$ are exactly the same as those obtained by simply truncating $d_n = 0$ for $|n| \geq 2$.

\section*{References}
\bibliographystyle{unsrt}
\bibliography{Stellarator-Secondaries}

\begin{thebibliography}{10}

\bibitem{diamond-zonal}
P.~H. Diamond, S.-I. Itoh, K.~Itoh, and T.~S. Hahm.
\newblock Zonal flows in plasma--a review.
\newblock {\em Plasma Phys. Control. Fusion}, 47(5):R35, 2005.

\bibitem{sugama-2006}
H.~Sugama and T.-H. Watanabe.
\newblock Collisionless damping of zonal flows in helical systems.
\newblock {\em Phys. Plasmas}, 13(1):--, 2006.

\bibitem{sugama-2008}
H.~Sugama and T.-H. Watanabe.
\newblock Turbulence-driven zonal flows in helical systems with radial electric
  fieldsa).
\newblock {\em Phys. Plasmas}, 16(5):--, 2009.

\bibitem{mishchenko}
A.~Mishchenko, P.~Helander, and A.~{K{\"o}nies}.
\newblock Collisionless dynamics of zonal flows in stellarator geometry.
\newblock {\em Phys. Plasmas}, 15(7):--, 2008.

\bibitem{helander-mischenko}
P.~Helander, A.~Mishchenko, R.~Kleiber, and P.~Xanthopoulos.
\newblock Oscillations of zonal flows in stellarators.
\newblock {\em Plasma Phys Contr F}, 53(5):054006, 2011.

\bibitem{rogers-prl}
B.~N. Rogers, W.~Dorland, and M.~Kotschenreuther.
\newblock Generation and stability of zonal flows in ion-temperature-gradient
  mode turbulence.
\newblock {\em Phys. Rev. Lett.}, 85(25):5336--5339, Dec 2000.

\bibitem{strintzi-jenko}
D.~Strintzi and F.~Jenko.
\newblock On the relation between secondary and modulational instabilities.
\newblock {\em Phys. Plasmas}, 14(4), 2007.

\bibitem{smolyakov-diamagnetic}
A.~I. Smolyakov, P.~H. Diamond, and M.~V. Medvedev.
\newblock Role of ion diamagnetic effects in the generation of large scale
  flows in toroidal ion temperature gradient mode turbulence.
\newblock {\em Phys. Plasmas}, 7(10):3987--3992, 2000.

\bibitem{anderson-pop}
J.~Anderson, H.~Nordman, R.~Singh, and J.~Weiland.
\newblock Zonal flow generation in ion temperature gradient mode turbulence.
\newblock {\em Phys. Plasmas}, 9(11):4500--4506, 2002.

\bibitem{anderson-jpp}
J.~ANDERSON and H.~NORDMAN.
\newblock The role of magnetic shear for zonal flow generation.
\newblock {\em J. Plasma Phys.}, 72(5):609Ð615, Oct 2006.

\bibitem{roberts-taylor}
K.~V. Roberts and J.~B. Taylor.
\newblock Gravitational resistive instability of an incompressible plasma in a
  sheared magnetic field.
\newblock {\em Phys. Fluids}, 8(2):315--322, 1965.

\bibitem{helander-review}
P.~Helander.
\newblock Theory of plasma confinement in non-axisymmetric magnetic fields.
\newblock {\em Reports on Progress in Physics}, 77(8):087001, 2014.

\bibitem{hasegawa-mima}
A.~Hasegawa and K.~Mima.
\newblock Pseudo-three-dimensional turbulence in magnetized nonuniform plasma.
\newblock {\em Phys. Fluids}, 21(1):87--92, 1978.

\bibitem{dorland-thesis}
W.~D. Dorland.
\newblock {\em Gyrofluid models of plasma turbulence}.
\newblock PhD thesis, Princeton University, 1993.

\bibitem{cohen}
Bruce~I. Cohen, Timothy~J. Williams, Andris~M. Dimits, and Jack~A. Byers.
\newblock Gyrokinetic simulations of {{\bf E}}$\times${{\bf B}} velocity-shear
  effects on ion-temperature-gradient modes.
\newblock {\em Phys. Fluids B}, 5(8):2967--2980, 1993.

\bibitem{plunk-itg}
G.~G. Plunk, P.~Helander, P.~Xanthopoulos, and J.~W. Connor.
\newblock Collisionless microinstabilities in stellarators. iii. the
  ion-temperature-gradient mode.
\newblock {\em Phys. Plasmas}, 21(3):--, 2014.

\bibitem{li-kishimoto}
Jiquan Li and Y.~Kishimoto.
\newblock Role of magnetic shear in large-scale structure formation in electron
  temperature gradient driven turbulence.
\newblock {\em Phys. Plasmas}, 12(5):--, 2005.

\bibitem{citrin-pop-2012}
J.~Citrin, C.~Bourdelle, P.~Cottier, D.~F. Escande, Ö.~D. Gürcan, D.~R.
  Hatch, G.~M.~D. Hogeweij, F.~Jenko, and M.~J. Pueschel.
\newblock Quasilinear transport modelling at low magnetic shear.
\newblock {\em Phys. Plasmas}, 19(6):--, 2012.

\bibitem{plunk-njp}
G.~G. Plunk, T.~Tatsuno, and W.~Dorland.
\newblock Considering fluctuation energy as a measure of gyrokinetic
  turbulence.
\newblock {\em New J. Phys.}, 14(10):103030, 2012.

\bibitem{dewar-glasser}
R.~L. Dewar and A.~H. Glasser.
\newblock Ballooning mode spectrum in general toroidal systems.
\newblock {\em Phys. Fluids}, 26(10):3038--3052, 1983.

\bibitem{plunk-pop-07}
G.~G. Plunk.
\newblock Gyrokinetic secondary instability theory for electron and ion
  temperature gradient driven turbulence.
\newblock {\em Phys. Plasmas}, 14(11):--, 2007.

\bibitem{jenko-dorland-pop}
F.~Jenko, W.~Dorland, M.~Kotschenreuther, and B.~N. Rogers.
\newblock Electron temperature gradient driven turbulence.
\newblock {\em Phys. Plasmas}, 7(5):1904--1910, 2000.

\bibitem{cowley-kulsrud}
S.~C. Cowley, R.~M. Kulsrud, and R.~Sudan.
\newblock Considerations of ion-temperature-gradient-driven turbulence.
\newblock {\em Phys. Fluids B}, 3(10):2767, 1991.

\end{thebibliography}

\end{document}